# Low Bending Loss Nodeless Hollow-core Anti-Resonant Fiber

Shoufei Gao, Yingying Wang*, Xiaolu Liu, Pu Wang

*National Center of Laser Technology, Institute of Laser Engineering, Beijing University of Technology, Beijing 100124, China*
*\* Email: wangyingying@bjut.edu.cn*

**Abstract:** We report a high performance nodeless hollow-core anti-resonant fiber with a broadband guidance from 1050 nm to >1900 nm, a transmission attenuation of 200 dB/km, and a low bending loss of 0.25 dB/m for a 5 cm radius at wavelength of 1550 nm.

Recent years have seen the advances in hollow-core anti-resonant fiber (HC-ARF) with a typical structure of one ring cladding containing several capillaries to form a negative curvature (NC) core-surround (also called NC-HCF) [1]. Though the guidance mechanism of HC-ARF is slightly different with the inhibited-coupling hypocycloid-core Kagome hollow-core fiber (Kagome-HCF) [2], the performance is similar, i.e. large pitch and large core size, low transmission loss, broad transmission band, high laser damage threshold, and mainly single mode guidance when keeping straight. Attributing to these outstanding performances, both types of fibers are now widely used in gas-based nonlinear optics, ultra-intense pulse delivery/compression, interferometric sensing, quantum optics, fiber-aided laser micromachining and mid-infrared or terahertz propagation. On the other hand, all the practical applications require a more flexible and convenient fiber that can be mechanically and optically bent to a relatively small radius. Regretfully, the performance of above two types of HCF degrades dramatically after being bent. Various approaches have been attempted to reduce the macro-bending loss. In Kagome-HCF, it was shown that by increasing the number of cladding layers from 3 to 4, the bending loss is greatly reduced to ~0.2 dB/m with a bending radius of 5 cm and wavelength of 1500 nm-1700 nm [3] (written as r=5 cm, $\lambda$=1500-1700 nm, same below). Though simulation results did not collaborated well with the experiment, it was qualitatively concluded that adding more layers of rings helps to reduce bending loss. However this implies a more complicated structure and a very large outer diameter toward ~ 300 μm. In NC-HCF, in order to achieve low bending loss, a more intuitive approach was proposed by Belardi *et al* via using a nodeless cladding structure to form a free core boundary [4]. Due to the absence of Fano resonance, the coupling strength between the core mode and the cladding mode is greatly lowered even in a bent state. Though a few nodeless fiber structures have been experimentally realized [4-6], their optical performance is not satisfactory due to either a thick strut [4, 5] or a small capillary size [6]. The current record of bending loss in HC-ARF is at the level of 1 dB/m with r=5 cm, $\lambda$=3100 nm [4] and 2.6 dB/m with r=6 cm, $\lambda$=1064 nm [1]. The tough requirement of thin strut thickness (wide transmission bandwidth), large capillary size (enough separation between hollow-core and silica jacket) and nodeless structure (low bending loss) has yet been realized in HC-ARF fabrication.

In this paper, we present, what we believe the first, high performance nodeless HC-ARF operating in the first transmission band with broadband guidance (from 1050 nm to >1900 nm), low transmission loss (~200 dB/km) and single mode guidance even in a bent state. More importantly, it's macro bending loss value of 0.25 dB/m with r=5 cm, $\lambda$=1500 nm is comparable to that of the more complicated four-ring Kagome HCF.

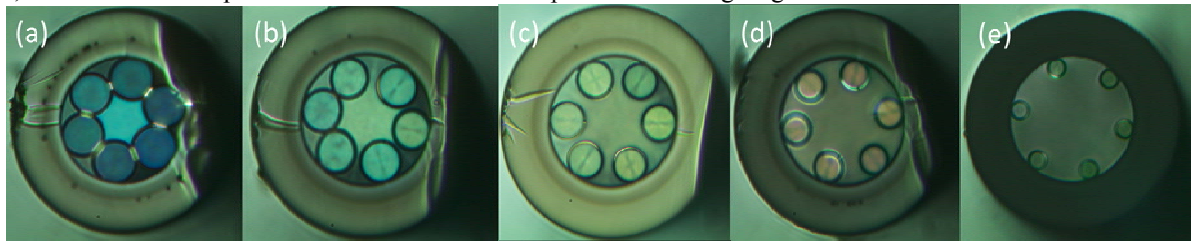

Fig. 1. Optical microscope images of a series of HC-ARF with (a) touched negative curvature core surround; (b) partly touched core surround; (c-e) free boundary core surrounds with varied capillary size.

Figure 1 shows a series of NC-HCF with outer diameter of 145 μm drawn from similar canes with six untouched capillaries. Finely adjusting the pressure inside the six capillaries, we are able to control their sizes to form a touched negative curvature core shape [Fig. 1(a)], partly touched core shape [Fig. 1(b)] and free boundary core shape with varied separation distances between adjacent capillaries [Fig. 1(c-e)]. For the structures of Figs.1(a&b), the optical spectra contain a lot of oscillations and are very sensitive to bending because of the Fano resonance. Figure 1(c) is our target structure with a nearly touched nodeless structure to perform as a delicate waveguide. However, in the drawing process, small variation of pressure may change the shape to Fig. 1(b) and thus only short length of fiber of Fig. 1(c) is obtained for now (20 m). We plan to increase draw tension to solve this stability problem. For the structure of Fig. 1(d), we have successfully drawn >100 m length. The measured optical performance is close to that of Fig. 1 (c). In Fig. 1(e), because of the shrink of the capillaries, the strut thickness becomes too thick to perform as a good anti-resonant waveguide. Thus, in this paper, we concentrate on the structure of Fig. 1(d).

Figure 2(a) shows the scanning electronic microscope (SEM) image of the fiber of Fig. 1(d). The six untouched capillaries with a diameter of 20 μm or center-to-center distance of 36 μm and a strut thickness of 410 nm form a free boundary core-surround with an inscribed radius of 46 μm. Thanks to the thin strut thickness, the fiber operates in the first transmission band similar to the Kagome-HCF. This means a much wider transmission bandwidth than the state-of-art NC-HCF that usually operates in the second transmission band. Figure 2(b) shows the attenuation spectrum measured by a cut-back method from 110 m to 10 m, exhibiting a transmission band from 1050 nm to >1900 nm (limited by our supercontinuum source) and an attenuation of 200 dB/km. This attenuation value, though higher than that of NC-HCF and Kagome-HCF, is acceptable for most applications. A larger capillary size and hence less separation between capillaries, i.e. the structure of Fig. 1(c), may lower this loss figure. Our bending loss measurement is performed by bending a 10m fiber to radii of 2 cm, 3 cm, 4 cm and 5 cm with 3, 8, 10, 10 turns simultaneously [Fig. 2(e)]. The loops are then released in sequence and the spectrum is recorded as shown in Fig. 2(c). The bending loss spectrum is hereby derived [Fig. 2(d)], showing a high loss in the short wavelength part especially for small bending radius. We plot in Fig. 2(g) the evolution of the bending loss value in two typical wavelength of 1550 nm and 1800 nm respectively, showing a low bending loss of 0.2-0.25 dB/m at r=5cm. At 1800 nm, the bending loss is at the level of 1dB/m even with r=3 cm. Furthermore, in Fig. 2(f) we show the single modeness is maintained in fiber with r=3 cm, λ=1500 nm.

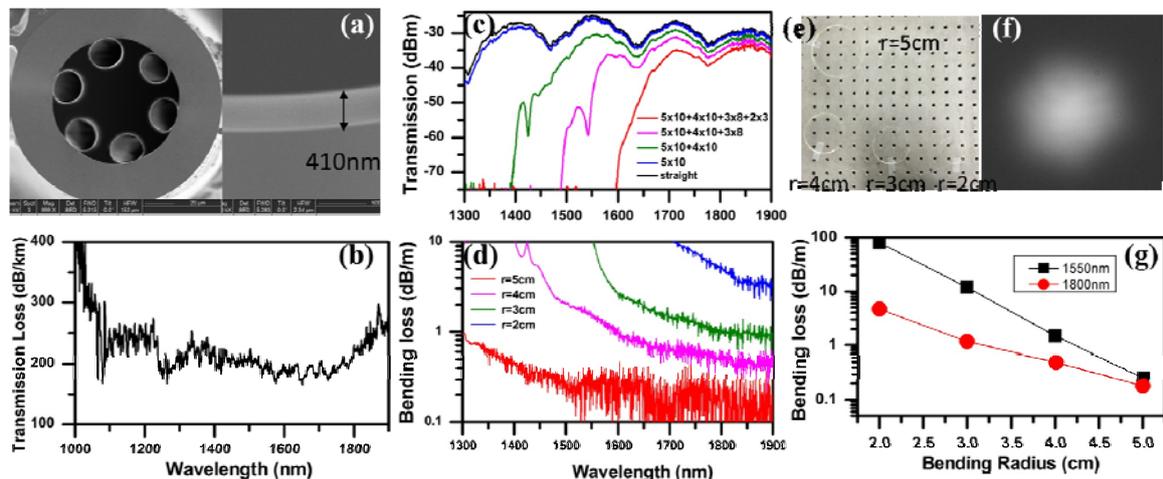

Fig. 2 Characterization of the nodeless HC-ARF in Fig.1(d). (a) SEM image of the fiber. (b) Transmission loss spectrum measured from 110 m fiber cut to 10 m fiber. (c) Transmission spectrum after bending. The legend in the graph is written in the form of *bending radius in cm* x *number of turns*. e.g. the red curve represents the spectrum after bending the fiber to 5 cm x 10 turns + 4 cm x 10 turns + 3cm x 8 turns + 2 cm x 3 turns as indicated in (e). (d). Bending loss spectrum derived from (c). (f) Far-field profile of the fiber with r=3 cm, λ=1500 nm. (g). Evolution of the bending loss value at two wavelengths of 1550 nm and 1800 nm.

In conclusion, we report a high performance nodeless HC-ARF with six untouched thin wall capillaries forming a free boundary core-surround. Realization of this challenging HC-ARF structure possesses three folded merits: 1) the nodeless cladding helps the fiber to maintain low loss for the core mode even at a bent state; 2) the thin capillary wall gives rise to a light guidance within a broad bandwidth of > 120 THz; 3) the single modeness of the core mode is attained by a suitable capillary diameter for resonantly out-coupling higher order modes.